# Lightweight Call Signaling and Peer-to-Peer Control of WebRTC Video Conferencing

Kundan Singh
Intencity Cloud Technologies
San Francisco, CA, USA
kundan10@gmail.com

## ABSTRACT

We present the software architecture and implementation of our web-based multiparty video conference application. It does not use a media server. For call signaling, it either piggybacks on existing push notifications via a lightweight notification server, or utilizes email messages to further remove that server dependency. For conference control and data storage, it creates a peer-to-peer network of the clients participating in the call. Our prototype client web app can be installed as a browser extension, or a progressive web app on desktop and mobile. It uses WebRTC data channels and media streams for the control and media paths in implementing a full featured video conferencing with audio, video, text and screen sharing. The challenges faced and the techniques used in creating our lightweight or serverless system are useful to other low-end WebRTC applications that intend to save cost on server maintenance or paid subscriptions for multiparty video calls.

## 1. INTRODUCTION

WebRTC or the web real-time communication [1][2] is a set of standard APIs (application programming interfaces) and protocols to enable low latency media paths between web browser instances. These browser clients need a *notification server* to exchange certain signaling messages between the clients [3][4]. It optionally needs a media relay, a media gateway and/or a *media server* for distributed media paths of a typical multiparty audio/video conference.

A significant portion of the cost of such systems often goes to the operation of these servers hosted on cloud/datacenter. Moreover, the capital expense of building such servers is also high due to their complex application logic. The complexity is largely due to proprietary vendor-specific implementation to tightly control the user access, signaling messages, and media servers in the conference. What if we could completely avoid these heavy-duty application and media servers, and the associated cost?

We describe an endpoint driven software architecture and implementation that enables multiparty video calls with little or no server dependency. It uses freely available mobile and web push notifications for call signaling to establish a WebRTC data channel between two endpoints. Subsequent signaling such as for media path negotiation is done over this data channel. As more users get invited to the call, it creates a peer-to-peer (P2P) network of clients participating using pairwise data channels. It uses this P2P network for storage and synchronization of conference data and the associated conference application logic entirely in the endpoints.

Our proof-of-concept implementation [5] of the client web app, named *Ezcall*, shown in Fig. 1, can be installed as a browser extension or a progressive web app (PWA). By default, our app uses a lightweight notification server with a couple of hundred lines of PHP code. This mode uses Google's Firebase Cloud Messaging (FCM) [6][7] for push notifications. Optionally, the user can remove this service dependency in the client app by using email or instant messages for call signaling. In that mode, the initial data channel setup requires sending a message and receiving a response, each with some call signaling data.

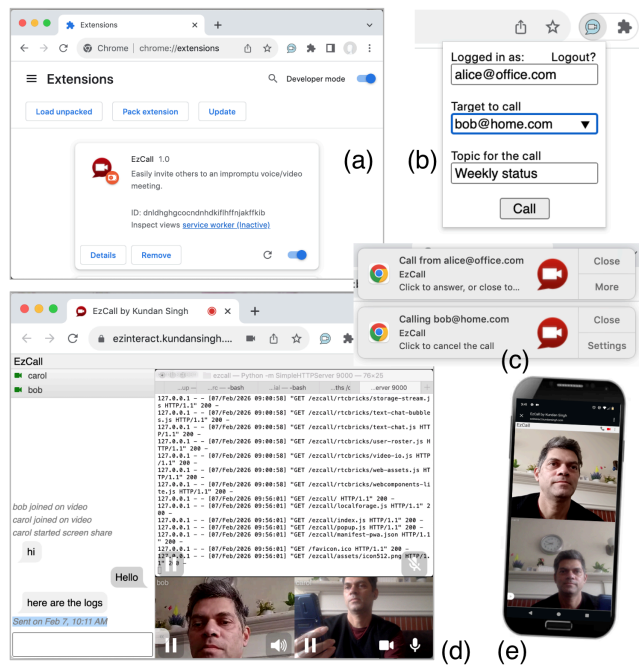

**Fig. 1** Screenshots of Ezcall: (a) the browser extension page in Chrome showing the app; (b) popup to send a call invitation when the extension icon is clicked; (c) notifications for incoming and outgoing call invites; (d) video conference web app with audio, video, screen/app share, and text chat in the wrapper app loaded in a browser; (d) video call on an installed mobile app (PWA).

Our implementation is in pure JavaScript without any client side framework. It uses either push notifications or email messages for the initial call setup. It does not use any other persistent service to operate, beyond the initial download and install of the web app from a static website, and optionally, a lightweight notification server.

The paper is structured as follows. Section 2 presents the background and related work. We describe the architecture and implementation details in depth in Section 3, and discuss the design decisions, security aspects, other challenges, and useful techniques in Section 4. We present the conclusions and ideas for future work in Section 5.

## 2. BACKGROUND AND RELATED WORK

A typical video call or conference involves three steps: lookup, session negotiation, and maintenance (Fig. 2). First, in lookup, a caller reaches a registered callee with a call invitation, which the callee answers to join the call; or an organizer sets up a meeting and informs the participants, e.g., via calendar invite to join, or to open the same web page to be connected to each other. This step requires the users to be reachable by some means, and/or be connected to some application on the network.

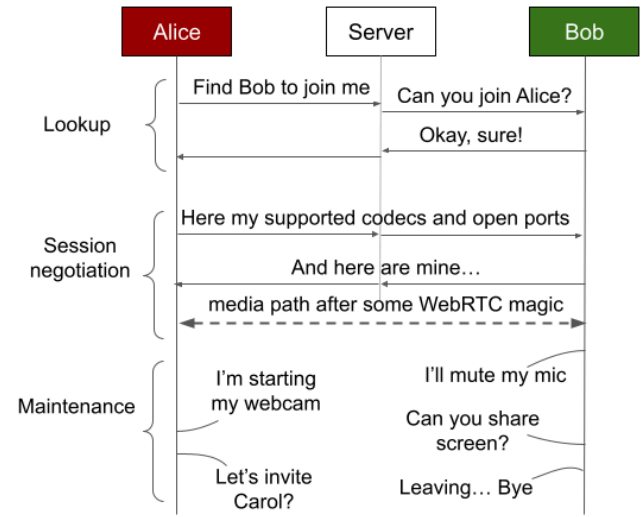

**Fig. 2** High level steps in a video call based on WebRTC.

Second, in session negotiation, the two endpoints, or one endpoint and the media server establish a media path by exchanging signaling messages. Multiple messages may be exchanged for SDP (session description protocol) offer-answer, and ICE (interactive connectivity establishment) candidates in trickle mode [2]. However, a minimum of two messages, one in each direction, is required, and is achieved by not using trickle ICE in offer-answer.

The first two steps may be combined as in the traditional SIP/VoIP (Session Initiation Protocol/Voice over IP)-based call flows. The third step is control and maintenance of ongoing call and conference, e.g. to notify the endpoints when participants join or leave, or adding or removing media channels such as for text or content sharing.

WebRTC [1][2] deals with only the second step above, and intentionally leaves out the first step from the protocol specification. Existing web video conferencing apps employ a wide range of techniques; some use SIP [8] or other open protocols, but many use proprietary mechanisms for a tight control of the application logic. Moreover, push notification or cloud messaging [6] is often used on mobile apps for initial call notification, which causes the app to wake up and connect to the application server for further steps.

We extend this to use push notifications in both directions to establish the initial data channel of the control path, and on both mobile and desktop. Google's FCM has a wrapper for push notifications on multiple mobile and desktop platforms, including Android, iOS and web. It uses a web service worker [7], which is a background process in the browser linked to a website, whose lifecycle is controlled by the browser, but is automatically activated when a push notification is received. It is also available for browser extensions and installed apps (PWAs). A PWA [9] is a web app that is designed to be installable on the desktop or mobile, can operate in offline mode if needed, and runs and behaves as a native app.

Video conferencing usually needs servers, not just for call signaling negotiation, but also in the media path [3]. A media relay enables media paths for clients behind a restricted firewall or NAT (network address translator). A media server facilitates media distribution in a multiparty conference, and can provide other features such as call recording or transcription. A media relay is usually not needed when a media server on the public Internet is used in the media path, and the server supports symmetric RTP and ICE lite. A media server can behave either as a gateway to transcode or interoperate with other systems, or an SFU (selective forwarding unit) by distributing a client's media to one or more other clients, or as an MCU (multipoint control unit) by mixing multiple clients' media before sending to each, or as a combination, e.g., forwarding video but mixing audio [10][11][12].

A wide range of media path topologies are possible using the SFU and MCU functions that can scale from two or three party calls to very large conferences. Furthermore, instead of the dedicated or hosted media servers, the web apps running in the browser can act as SFU or MCU, for small scale conferences [13][14]. These techniques are directly applicable to our system.

Endpoint driven video conferencing systems implement the conference and call control application logic in the endpoint

2 © 2026, Kundan Singh

and use real-time database services for structured data storage, and notification of data changes [15][16][17]. We extend this idea further by avoiding a server, and by implementing such a data storage on an unstructured P2P network of participating clients.

In summary, all the pieces of our project including push notification/cloud messaging for call signaling [6][18][19], email or instant message for session negotiation [20][21], endpoint driven call/conference application logic separate from its data storage [16][17][22], P2P network for call and conference control [29][30], and serverless media paths in web apps [13][14] have already been independently proposed or researched. The depth of our effort is in combining all these techniques to create a compelling software architecture and a practical implementation.

## 3. SOFTWARE ARCHITECTURE AND IMPLEMENTATION

We start with a high-level block diagram of the endpoint software. Then we detail the notification service including its database schema and API. This is followed by the details of push notification based call signaling, including example message flows, and state machine. We also show how the serverless call signaling works using emails. Then we cover the multiparty conference logistics, including the shared data model, and audio/video path setup. Then we present how a peer-to-peer (P2P) network is formed among the clients participating in the call. Finally, we explore various media path topologies including full mesh, centralized and hybrid for real-time media distribution in a conference.

### 3.1 Block Diagram of the Endpoint Software

Fig. 3 shows the layered architecture of the software in the endpoint. Ezcall is a web app that can be opened as a web page, or installed as a browser extension or an app (PWA). For signaling, it uses either push notifications, or email/instant messages sent and received by the end user. For the former, it uses FCM, as well as a lightweight notification server, named contacts.php, as shown below.

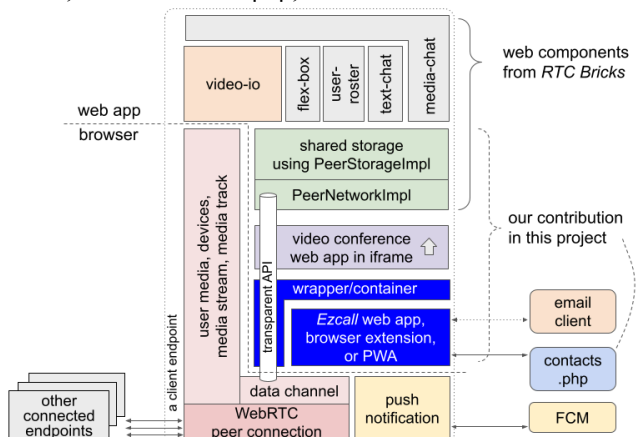

**Fig. 3** Block diagram of various layered components in the endpoint software. Also indicates this project's contribution.

During call signaling, it launches a new browser tab, with a web application named wrapper/container, to negotiate and establish a WebRTC peer connection and a data channel with a remote endpoint. In a multiparty call, the wrapper manages multiple data channels, one for each connected remote endpoint or peer.

The wrapper also negotiates and loads one or more web apps in iframes in that browser tab. It provides them with a shim API layer to send or receive messages with other endpoints on the data channels. The negotiated app can be configured by the user. By default, the wrapper loads a video conference web app, media-chat-peer.html, that we built using the web components from the *RTC Bricks* project [22]. That project contains several web components useful for building real-time multimedia communication and collaboration applications. It uses a resource-based software architecture to separate the application logic from its data storage service, while keeping the application logic in the endpoint [16][17].

We replaced its default and centralized data storage, with our new P2P network, and an associated distributed data storage. We built two components, PeerNetworkImpl and PeerStorageImpl, where the former implements an unstructured P2P network node, and the latter uses that to implement a shared data storage. We preserved the existing API of the shared-storage component, so that all the existing web components of that project continue to work in our P2P storage. The media-chat-peer and wrapper collude to expose a WebRTC data channel like API to PeerNetworkImpl. This makes our new components readily reusable in other projects that use data channels, beyond this Ezcall use case.

### 3.2 (Optional) Notification Service

Although user signup is optional in our application, we keep it outside the scope of Ezcall. Any existing mechanism may be used to sign up, by creating a mapping from an auth token to an user identifier at the notification server. By default, we use email addresses as user identifiers. The auth token is a randomly assigned UUID (universally unique identifier) that remains fixed for a single installation. One user identifier may be associated with multiple tokens, and one token identifies one installation of the app. Additionally, an FCM token obtained by the client app from the FCM service identifies the installed app for the purpose of push notifications.

*Database schema*

The notification server keeps track of the registered user's identifier, auth token, and FCM token using the following database schema. Although the user identifier is named email, the value can be anything that the user prefers to be identified by others such as a phone number, social media profile link, or fingerprint of the user's public key.

| Field | Type | Description |
|---|---|---|
| token | text (key) | Randomly assigned UUID at signup representing an app install |
| email | text | Email address or other value for user identification |
| created | date/time | To keep a record of external signups |
| contact | text (uniq) | The currently active FCM token, linked to an install or endpoint |
| expires | date/time | To auto refresh the token to contact mapping by the endpoint |

*Client-Server API*

Our notification server exposes a web API to update the mapping, and to send a push message to the target user's FCM tokens given her user identifier. To login, the client sends the auth token in the HTTP Authorization header, and receives the associated email address on success, as shown below.

```
=> GET /contacts
   Authorization: Bearer <auth token>…
<= { "email": "alice@office.com" }
```

This API just verifies that the auth token exists, and is associated with a registered user identifier.

An FCM token identifies the device or app that can receive push notification from the FCM service, and represents the user's contact or reachability. There may be only one contact for each auth token or app installation. To register the FCM token, the client sends an HTTP PUT request, with the body as shown below.

```
=> PUT /contacts
   Authorization: Bearer <auth token>…
   { "contact": "FCM token…" }
<= { "instance": "abc12", "expires": … }
```

The server updates the FCM token in the database for the associated auth token of the request. Thus, a new FCM token for the same auth token overrides the previous. The response includes an unique instance identifier for this contact, which may be used by the sender to target a specific contact, versus all contacts. Internally, the server calculates the instance identifier using a simple 32-bits CRC (cyclic redundancy check) checksum on the combined value of the user identifier, auth token and FCM token.

When a client wants to send a push notification to a target user, it uses an HTTP POST request with a body containing the target user identifier, as shown below.

```
=> POST /contacts
   Authorization: Bearer <auth token>…
   { "to": "bob@home.com", "data": {...} }
<= { "count": 2 }
```

The target, in the "to" attribute, may be a user identifier shown above, or a specific contact device with an instance identifier, e.g., "bob@home.com/abc12". On success, the response contains the count of how many contacts of the target were successfully used to send the push notification. In case of failure, a 404 response with a body containing additional details is returned. Note that if multiple contacts are present, then successfully sending to any contact results in a successful response.

When the notification server receives this request, it verifies the sender's auth token, and that the "data" attribute is an object. It then locates all the contacts of the specific target from the database, and attempts to send the data value using the FCM push notification API for each contact. It adds or replaces the "To" and "From" attributes in the data object, to indicate the specific contact of the target and the sender, respectively, in each push attempt as shown below. This helps the receiver of the push notification to verify if it is the right contact to receive the message, and for it to send a response to the specific contact of the sender.

```
=> POST https://fcm.googleapis.com…
   {"message":{"data":{...,
     "To":"bob@home.com/abc12",
     "From":"alice@office.com/xyz34" } }
```

By default, our push notification expires after a minute, as it is used for real-time call signaling only. We use oAuth based FCM v1 API, instead of the deprecated legacy one.

### 3.3 Call Signaling Flow with Push

When the Ezcall app is started, it shows a screen (see Fig. 4) with two options: login to use push notification for call signaling by entering the auth token, or use email messages for call signaling in a serverless mode.

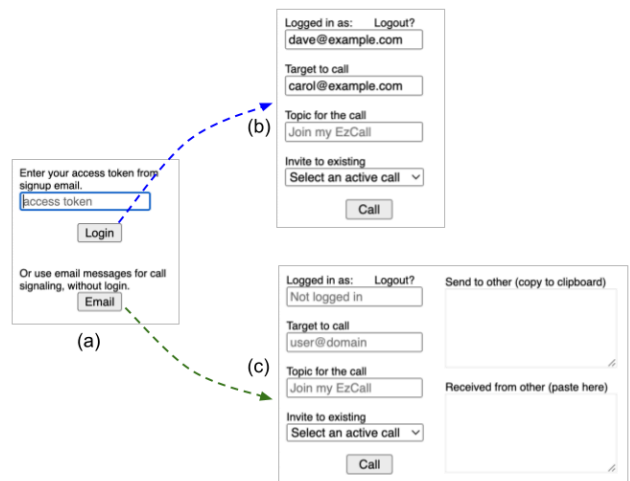

**Fig. 4** (a) Login (or email) option; (b) after login, call using push notification; or (c) if no login, send or receive call invite using email messages - it allows copying and pasting the messages.

If the user chooses to login, the client initiates the login API mentioned earlier, and saves the auth token in the browser's local storage to avoid future prompts about login. It then retrieves the FCM token (step 1, in Fig. 5), and registers that token at the notification server (step 2).

*Sending an outbound call invite*

Fig. 5 shows the call signaling to establish a bidirectional data channel between two endpoints. In this example, when Alice initiates a call to Bob, the app creates a new random *invite identifier*. It shows a notification for this outgoing invite, so that the user can interact with it, e.g., to cancel the pending invite anytime, until the invite is answered or times out. The app then launches a new browser tab with the wrapper app (step 3). The app picks a randomly generated *node identifier*, and supplies it to the wrapper. This node identifier is used throughout the call, and the P2P network, to identify this client endpoint. It must be unique across all the clients within a single conference. The app also creates a new random meeting or *conference identifier*, and instructs the wrapper to load the video conference web app using this conference identifier. Note that the web app includes a PeerNetworkImpl object, which represents a node in the P2P network, with the node identifier previously selected. Finally, the app instructs the wrapper to create a peer connection and a session "offer" for a new data channel with this invite identifier.

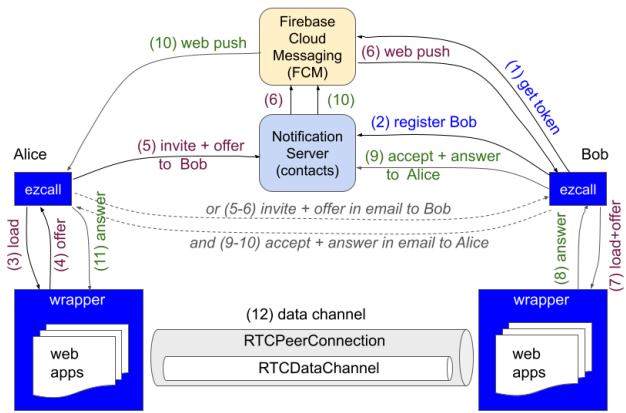

**Fig. 5** Call signaling steps to establish a data channel.

The wrapper uses non-trickle mode to generate a single offer SDP containing the ICE candidates for a data channel, and gives it to the app (step 4). The message includes the invite identifier so that the app can co-relate the offer with the pending outgoing invite, and can support multiple simultaneous call invites. The app then constructs an invite message containing all the details of the call including the invite identifier, the conference identifier, the sender's node identifier, an optional topic of the call, and the offer SDP.

If the caller, Alice, is logged in, the client uses the push notification to send the invite message via the notification server to the target user, Bob (steps 5-6). The push notification wakes up Bob's client's service worker on the receiver side, which shows a notification to the user (see Fig. 1(c)). The receiving user, Bob, may interact with the notification to answer or decline the call invitation. In case of no interaction after a minute, the app times out and removes the call invite notification.

*Accepting or declining an inbound call invite*

If the user declines the call, a new decline push notification, with necessary data such as invite identifier, is sent back to the caller, using the "From" value of the notification message. Thus, it goes back to the specific instance of the caller's app that sent the invite.

If the receiving user, Bob, answers the call, the app attempts to create the receiving side of the data channel. Note that the WebRTC API for creating a peer connection or data channel does not work from the service worker. Hence, if the user accepts the call invitation by clicking on the notification, and the main Ezcall app window is not already open, the service worker first opens the main app window. It then launches a new browser tab, with the wrapper and supplies the necessary information from the call invite, including the conference identifier, invite identifier, node identifier, optional topic, and the offer SDP (step 7) .

Similar to the caller side, the receiver app also generates a new unique random node identifier to assign to each wrapper. The app also instructs the wrapper to load the video conference web app using the conference identifier, and to create a peer connection and session "answer" for the received "offer" of the data channel in this invite. The wrapper uses non-trickle mode, the same as the caller side, to generate a single SDP containing the ICE candidates for the data channel, and gives it to the app (step 8). This message includes the invite identifier for co-relation at the caller side.

The app then constructs an accept message containing all the details including the invite identifier, and the generated answer SDP, and sends it back to the caller's specific contact based on the "From" value of the initial notification message. This is sent using another push notification from the receiver to the caller (step 9-10). The caller's app receives the message via push notification, updates the existing notification display indicating that the target user has accepted the call, and delivers this answer SDP to the wrapper (step 11). This in turn completes the offer-answer negotiation of the peer connection at the caller side. The callee side negotiation completed earlier (after step 8).

If the offer-answer negotiation is successful, and the wrapper apps at the two endpoints are able to find a network path to establish a WebRTC data channel, then both the sides declare successful connection (step 12). Then they use this data channel for subsequent data exchanges, e.g., for actual audio/video media path establishment and conference control. These are described later.

Fig. 6 shows the complete state machine for each invite at the endpoint during call signaling. Note that the declining and active states are almost similar, and both transition to idle after a short timeout. Since this state machine only applies to the call invitation, it stays active only during the initial call setup phase, and not for the duration of the call, similar to an INVITE transaction at a SIP proxy [23].

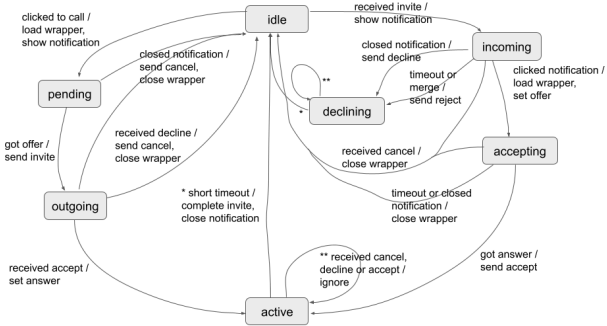

**Fig. 6** State machine for call invite at the endpoint.

*Merging a received invite to an existing call*
If the receiving user, Bob, is already in a call when an incoming invite is received, then he may decide to answer it as a new call (described previous), or merge it with the existing call. If Bob decides to merge, then a special call rejection response is sent back to the caller, indicating that the receiver will call back with a new invite. In that case, Bob's app then follows the message flow from the caller side to invite Alice's specific contact, in the existing call, using a new call invite and a new state machine, but using the same conference identifier as that of Bob's existing call.

From the user experience perspective, the merge option may be automatically selected, if an active call is focussed in the browser when the call is answered. Similarly, when an outbound call invite is done while an active call is focussed in the browser, then the call invite is automatically assigned to the same conference identifier, to invite the new user in the existing conference. Alternatively, the user may use specific notification buttons to respond (Fig. 1(c)).

*Forking call invite to multiple devices*
The app supports multiple logins by the same user identifier. This is loosely based on the SIP forking proxy behavior [23] as shown in Fig. 7. It shows the caller, Alice, on two devices, (x and y), and the receiver, Bob, on three: (a, b, and c).

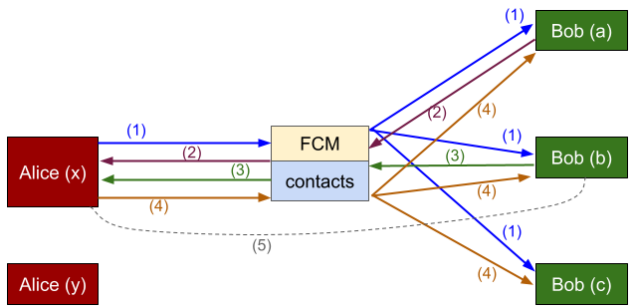

**Fig. 7** Call signaling steps to establish a data channel, when the two users have multiple logged in devices.

The forking of messages is done by the notification server as follows. If it detects multiple contacts for the target user identifier, it generates multiple push notifications, one per contact. Thus, Alice's call invite to Bob is forked to all the three contacts (flow 1, in Fig. 7). In this example, the first device, (a), times out, and sends a reject message (flow 2), which is ignored at the caller's end. The second device, (b), is used by the receiving user, Bob, to accept the call invitation (flow 3). This causes the caller side to send a cancel request (flow 4), so that all the other pending devices can cancel the incoming invite. Once the accept is delivered to the caller, a WebRTC data channel is established between the two specific devices (flow 5) using the offer and answer negotiation done earlier (flows 1 and 3).

The decision to send the cancel request can be based on the count of the push notifications sent, as indicated in the response from the server (after flow 1). But for simplicity and robustness, our implementation always sends a cancel request when the call is answered, or declined. The receiving client that actually answered the call (or rejected, or timedout) simply ignores the received cancel request as shown in the state machine. The other pending receiver clients can use that to remove the call invite notification.

Note that we treat the explicit call decline by the end user differently than other reject reasons such as timeout, busy, or merge. In particular, a decline causes immediate cancellation of the invite from the caller side, whereas other reject reasons let the caller or end user decide whether to cancel the call invite or not. Moreover, a missed call notification is not shown for declined calls.

Note that in Fig. 7, messages from the receiver to caller direction always use the specific contact of the caller device, hence are only delivered to the specific client that sent the initial invite (e.g., flows 2 and 3). On the other hand, the messages from the caller to receiver direction, including the initial invite (flow 1) or the later cancel (flow 4), are sent to the receiving user's identifier by default, which are forked to all the contacts of that user.

### 3.4 Call Signaling Flow with Email

The serverless call signaling flow behaves almost similar to the previous call signaling flow with push notifications. There are some differences, shown in Fig. 5, and described below.

In the previously described call signaling flow, after step 4, if the caller is not logged in, the Ezcall app opens the user's email client using the "mailto:" URL. The to, subject and body are prepopulated with the necessary information to send the invite via email. The user is instructed to send the email to the target, Bob (step 5-6). The invite data is also shown in the app's send box (Fig. 4(c)), so that Alice can send via some other instant messenger. When the user, Bob, receives the invitation via email, he launches his Ezcall app if not already open, and pastes the email body to the appropriate received message area of the app (see Fig. 4(c)). This triggers an incoming invite in the app. The rest of the process, such as loading the wrapper, creating various identifiers, or showing the notification about outbound or incoming calls, is the same as before.

© 2026, Kundan Singh

When the user clicks on the incoming call notification, to accept the invite, and the app receives the answer SDP from the wrapper (step 8, in Fig. 5), it opens up the user's email client using the "mailto:" URL. The body is prepopulated with the necessary information to send the response via email. As before, the app shows the accept data in the send box, so that Bob can send it using other messaging tools. The user is instructed to reply to the email from the caller, Alice (step 9-10). When the caller, Alice, receives the response via email, she copies and pastes the email body to the appropriate received message area of the app. This triggers the call accept in the app.

The app's user interface (see Fig. 4(c)) has two boxes - one to copy the message to send to the remote user, and the other to paste the message received from the remote user. Note that both the caller and callee will need to use both the boxes once for a successful call signaling - the caller sends the email first, and receives the response email; and the receiver receives the email first, and sends the response email to accept.

The email based flow has some other differences with push based flow, such as the use of additional identifiers to co-relate the responses with the requests. There is usually no cancel or decline flow, because a timeout can be used to expire the invite. This makes the state machine simpler for such cases. Moreover, there is no forking of call invitations. Since a call invite goes to the user's email, and not to the devices directly, the receiving user can explicitly pick the right device to paste the received invite to, and thus, to accept the call from.

## 3.5 The wrapper and the conference web app

As mentioned earlier, once a data channel is created between the two endpoints, in particular by the wrapper, the initial call setup is complete (Fig. 3 and 5). Afterwards, the actual multimedia call setup between the two endpoints is done by the video conference web app, media-chat-peer, loaded by the wrapper in an iframe of the same browser tab.

### Interface between wrapper and web app

The wrapper and the web app use cross-origin postMessage to communicate. This allows the wrapper to load an app from a different website or origin in its iframe if configured by the user. The wrapper passes some initial data to the web app in some URL parameters: data.path, data.self and data.displayname for identifying the conference resource, local user identifier and her screen name in the shared data storage; topic for subject of the call; and id identifying this node in the P2P network. Once the web app is loaded, it dispatches a "ready" message to the wrapper. Alternatively, the wrapper can listen for the "load" event on the iframe to know when the app is loaded.

Note that the wrapper maintains all the active WebRTC data channels, but those are used by the web app. When a data channel is set up between two endpoints, their wrapper apps first exchange their node identifiers. The wrapper informs the local web app about any data channel connection or disconnection, using an "add" or "remove" message, respectively, containing the other node's identifier.

The web app maintains the P2P network, and the shared data storage. It sends and receives messages on the active data channels to and from the other nodes. To send, it uses a "send" message with the target node's identifier and the actual message as a string. When the wrapper receives a message on a data channel, it delivers it to the web app using a "message" message, containing the sender node's identifier and the received message as a string.

### Peer-to-peer network and distributed data storage

The web app uses PeerNetworkImpl and PeerStorageImpl as implementations of a P2P network node, and the associated shared data storage, respectively. This data storage is shared among all the connected endpoints in the call (i.e., the nodes in the P2P network). All the endpoints view the same data in their storage. There may be a slight synchronization delay, or the data storage may get partitioned due to any data channel disconnection.

This distributed data storage is exposed as a shared-storage object to the other web components for various conference functions that are described next. Those web components, originally designed to work with a centralized data storage, can now transparently work with our P2P data storage, just by replacing their shared-storage reference.

## 3.6 Multiparty conference with media-chat

The actual multimedia call setup and conference control is done using the media-chat web component of the RTC bricks project. We go over the multiparty conference logistics of this web component below.

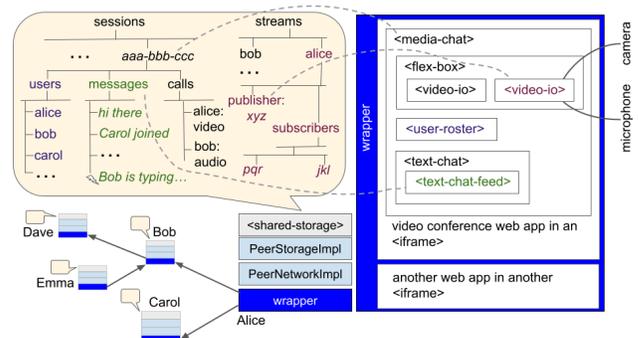

**Fig. 8** Web components of RTC Bricks, and their logical data structure, in a shared-storage for a multiparty conference.

The web components in that project use a shared-storage component to create and manage a hierarchical data structure, or resources, as shown in Fig. 8. An app can update a resource, and can receive notification on any change in real time when another app, on another endpoint, updates that resource.

A media-chat instance represents a user of the multiparty audio, video, and text conference, and binds to a session resource path, such as /sessions/aaa-bbb-ccc, where aaa-bbb-ccc is the randomly generated and unique meeting or conference identifier. This component embeds other components such as a flex-box for layout of participants' videos, an user-roster for displaying a list of participants, and their statuses, such as typing or active webcam, and a text-chat for muti-party text conversation. It provides a sub-resource path to these embedded components under its own resource path, e.g., text-chat binds to the sub-resource /messages, and stores all the text messages under that; and user-roster binds to /users to store the call participant data. The /calls sub-resource identifies which participants in the call have their audio or video enabled, so that the media-chat component can add a video-io component within its flex-box. The text-chat component uses the text-chat-feed component for display of chat messages, and a text input area for entering a message to send.

When the user enters a text message in the text-chat component's text input area, and hits enter, the component creates a new message data object under its resources, e.g., /sessions/aaa-bbb-ccc/messages/hya62. This object has the details of the message, e.g., {"sender": "Alice", "data": "hi there", …}. The text-chat components running in the app of the other connected endpoints receive a notification of this change, determine that a new message has arrived, and update their text-chat-feed components to display the new message from this sender. Some events such as typing indication is not created as a data resource, but sent as an explicit notification event on the resource path, and is delivered to the connected endpoints in the same way as a resource change event.

Similar resource manipulation and update notification is used by various web components to implement a full featured video conferencing experience. For full details on how these web components behave, interested readers are referred to the RTC bricks project [22].

## 3.7 Media Streams for Audio/Video

A video-io is responsible for publishing or playing a single audio or video media stream, interfaces with the camera and/or microphone devices of the endpoint, and is described in the next sub-section. It includes the user interface and application logic for publishing or playing a single media stream containing audio and/or video tracks using WebRTC. In our video conference app, it displays a single user's video: either the local user's published video, and a remote user's played video. It is also used for screen or app share.

The video-io component attaches to another named-stream component, which can be implemented in a variety of ways to support the session negotiation. We use a storage-stream implementation attached to the shared-storage component. It uses hierarchical resources representing zero or one publisher, and zero or more subscribers for each named stream.

Fig. 8 and 9 show that each user in the conference creates a named stream based on her identifier, e.g., /streams/alice. This user's video-io component for publishing binds to the /publisher sub-resource of the stream. The other users' video-io for viewing the stream, bind to a sub-resource under /subscribers. The WebRTC session setup using SDP offer-answer and ICE candidate exchange is done by sending notification events containing negotiation data on these sub-resources. Unlike the previous data channel, we use trickle ICE here. Depending on whether audio, video or both are used, the established media stream from a publisher video-io to a subscriber video-io has one or two media tracks.

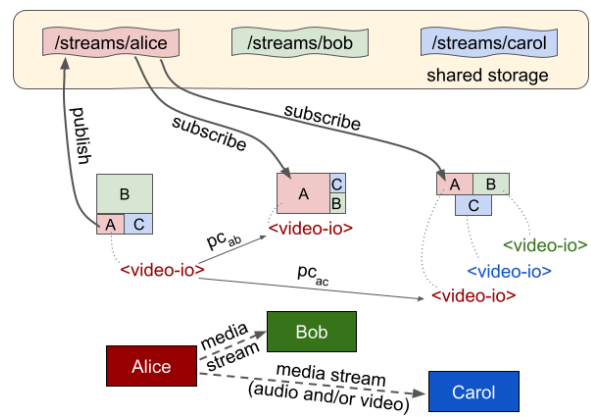

**Fig. 9** Interaction between and behavior of the video-io components and the named streams using shared-storage

The video-io component uses one peer connection and one media stream for each publisher-subscriber pair. Thus, in a two party call, there are two peer connections for media streams, one in each direction of the media path. Similarly, in a five-party video call, for media paths, there are N(N-1) =20 peer connections, sending 20 unidirectional media streams, one per connection, in a full mesh topology.

These connections are in addition to the initial peer connections during call setup for the control path. Although the number of peer connections can be optimized, and reused, we prefer to keep them separate. Separate peer connections for initial control path vs. subsequent media path allow the software to be modular, where the initial call setup does not deal with media path topology, or its optimizations, and the media path setup can happen independent of the control path.

## 3.8 Peer-to-Peer Network of Control Path

A P2P network is formed using the initial peer connection, and data channel setup, among the call participants. This is an unstructured or ad hoc network that mimics the call setup process. In particular, the network is based on who invited whom. Fig. 10 shows an example network with five users, where Alice invites Bob and Carol, Bob invites Dave, and Emma calls Bob who merges the received call with his existing call.

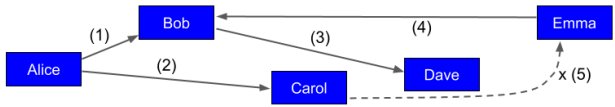

**Fig. 10** Peer-to-peer network with *control* path. Arrow is the call invite direction, and is a bidirectional data channel.

Because of the shared data storage, each node knows about the full topology of the control path, and if a cycle is detected, it gets resolved immediately. For example, if Carol attempts to invite Emma in the same call, with the same conference identifier as Emma's existing call, no new link is created in the network. However, if Carol invites Emma in a new call, using a new conference identifier, then that proceeds as a different call with its own different P2P network, separate from the previous call. We do not support merging separate calls with different conference identifiers currently.

We use flooding for data synchronization in the peer-to-peer network. Note that each link in the network is a bidirectional data channel. In general, a cycle is harmless, beyond causing redundancy in data transfer for synchronization of the shared data storage. For example, when PeerStorageImpl detects a change in the local data storage, it sends that change to all its immediate links. Each of the adjacent nodes further propagate the received message to the next nodes that they are connected to. A random and unique message identifier, and a small TTL (time-to-live) are used to ensure that the message is delivered or discarded after a while. This TTL is based on the number of hops similar to the IP packet's TTL, and defaults to a small number.

In addition to network flooding for data synchronization, the implementation also allows sending a message to a specific node, e.g., for session negotiation of audio/video streams, and private text messages. Each node maintains a simple routing table indicating the next hop the message should go to, for a given destination. The nodes synchronize the routing table with immediate neighbors, and update their own on any change.

Fig. 11 and 13 shows our test application to evaluate the data synchronization and routing on the P2P network. It can emulate up to 30 nodes for testing on a local machine. The network shown on the top-left is based on the call setup control path. The text messages, user roster and video call state in all the media-chat components are synchronized.

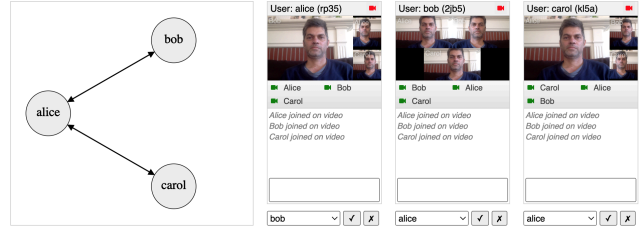

**Fig. 11** Test application showing a three party multimedia call.

Fig. 13 demonstrates a multiparty conference with text chat among six users, and shows how a broken link affects the conference, causing it to partition. When the partitions merge, the data resynchronizes. Message identifier is used to avoid duplicates after resynchronization. And in the future, message timestamps can be used to keep the order of messages the same for all users. The similar partition and merge behavior is observed for the audio/video chat as well.

## 3.9 Topology of Media Path

As mentioned earlier, the control path is independent of the media path, and the media path is full mesh by default. The media path is used only when the participating users enable audio/video in the multiparty chat.

Fig. 12 compares some of the media path topologies using application-level MCU and SFU functions (similar to [13]). The full mesh topology enables a low latency media path among the participants, and is useful for real-time conversation. However, network usage at each node scales linearly with the number of participants, i.e., each node receives N-1 and sends N-1 media streams. Uplink bandwidth can get congested easily for the users on low end networks. Moreover, separate streams may also put extra encoder processing load on the CPU/GPU of the machine.

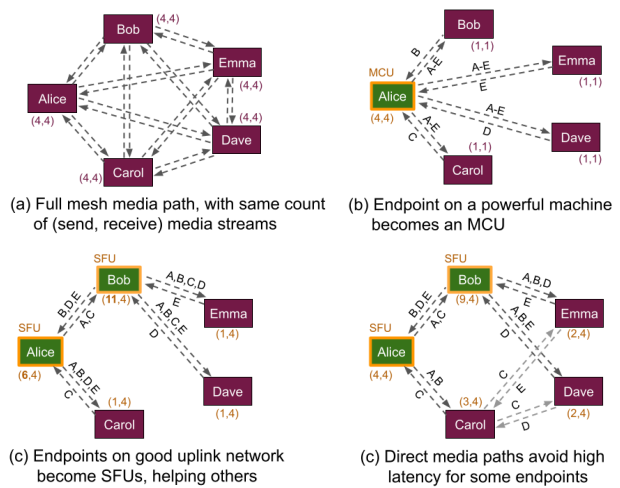

**Fig. 12** Various topologies of the *media* path. Arrow is the media stream direction. Each stream is on a separate peer connection.

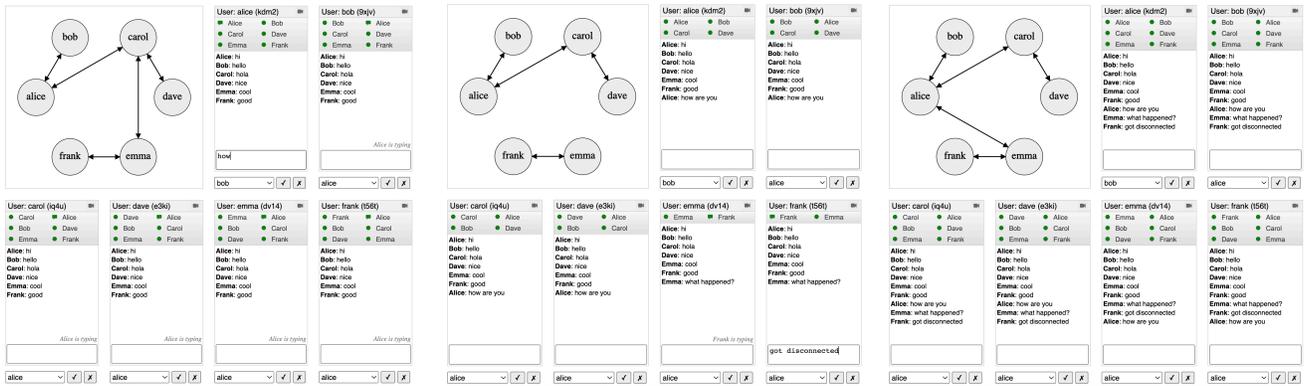

**(a)** Peer-to-peer network of control path shown on top-left. Also shows each participant's media-chat component.

**(b)** The link between Carol and Emma is broken. The conference partitions, and each partition continues separately.

**(c)** Alice invites Emma again, merging the two partitions. The data is resynchronized to show all the messages.

**Fig. 13** A test application to emulate up to 30 participants, create a peer-to-peer network of connection paths, and evaluate the message synchronization and media paths among them. Although running on the same machine, each participant's app behaves as an independent web app using the actual WebRTC connection, data channel and media stream APIs to form the control and media path.

One of the nodes that has a powerful CPU or processing capacity can act as a mixer, or multi-point control unit (MCU), as shown in Fig. 12(b). In that case, all the other nodes send the media stream to this one, and this one mixes all the received streams, generates a single outbound stream, and sends to each of the other nodes. This reduces the network as well as processing load on the other nodes, but causes a slightly higher processing load at the MCU node.

For video, the MCU draws a layout of various video elements (received in media stream from others, as well as the local preview), on a `canvas` element, and captures the stream from the `canvas` for a mixed video layout. For audio, to avoid listening to your own voice, the MCU should remove the speaker's audio from the mixed stream, before sending back the mixed stream to that speaker's endpoint. A drawback of using an MCU is that the latency is higher; not just due to the extra hop in the other users' media path via the MCU, but also due to the additional delay in playout and mixing at the MCU.

One or more of the nodes that are on a good uplink network can act as a forwarder, or selective forwarding unit (SFU), as shown in Fig. 12(c). This allows the other nodes to send only one stream on its uplink, to only one of the SFUs, whereas an SFU itself may send out more than N-1 streams. Although this reduces the uplink bandwidth for low end endpoints, it can cause significant increase in latency on the media path. Fig. 12(d) shows how some media streams may be obtained via a direct link instead of the SFU, to reduce the longer latency paths. Such a hybrid of SFU, MCU and full mesh media paths may be useful for very large endpoint driven conferences, where low latency is less important, such as webinars or event broadcasts.

Note that simulcast or SVC (scalable video coding) feature that is useful for server based SFU, is less useful in SFU nodes. Due to the desire to reduce the node's network usage, such SFU nodes should always use a single layer or level of video when sending out to the other nodes.

The decision to use an SFU or MCU, or to promote an endpoint to an SFU or MCU is done in a distributed manner. The decision can be based on heterogeneity of the endpoints, and the real-time quality and performance monitoring to detect poor network condition, or overloaded processing capacity at the endpoint.

## 4. FURTHER DISCUSSION

We discuss some non-trivial design choices, security considerations, technical challenges, and potential solutions in our implementation.

*Push notifications*

Silent web push notifications are no longer supported by the Chrome browser, i.e., it must always show or update a notification to the user. Since we use this for bi-directional offer-answer of the initial call signaling of the data channel, the app shows the notification at both the endpoints during call setup. Instead of showing new notifications, we update the existing notification message to reflect the call setup progress, e.g., when the call is answered, declined, or cancelled, so that only one notification is shown at any time for each invite in each direction. The notification persists and requires an user interaction for the initial call invite, and then auto hides after a timeout for subsequent events.

FCM triggers a token refresh event on Android, when the token is changed for any reason. However, such an event is no longer present for web push. In general, the token stays active for a very long time, of several months, but could be deleted or become inactive for reasons beyond the control of the client web app. In the absence of a guaranteed token refresh event, our web app's service worker attempts to get the FCM token every time it is started or activated. This enables the app to be reachable, even if the user has not opened the app for a long time. Additionally, the client app could use the Periodic Background Synchronization API available in the service worker, or the alarms API in the extension, to schedule a periodic refresh, say once a day, of the token. The user must grant this permission, in addition to existing camera, microphone, notifications, and pop-ups.

 © 2026, Kundan Singh

Since the control and media paths are separate, it is also possible to switch the call to a mixed topology, where control is peer-to-peer but media is centralized, or vice versa. For a centralized control path, the centralized resource server of the RTC bricks project may be used. For a centralized media path, external SFU or MCU servers can be deployed. Push notifications are still useful for the initial call invite notification even with a centralized control path.

Usually a server is needed to invoke the FCM API to send a push notification. Although a web app can invoke the API from JavaScript, it will require keeping the FCM authentication keys in the app, and hence, is not recommended. We show how a lightweight notification server can do this. Alternatively, a cloud hosted API gateway with lambda function may be used to actually send the push notification [24][25]. A free service [26] that can use Firebase or BitTorrent, among others, for a serverless call setup or matchmaking can also be used.

Finally, for a privately available app that is used only among trusted parties for communication, keeping the FCM keys in the endpoint app is not a problem. However, such systems still need to keep track of the mapping between user identifiers and FCM tokens. One could use publicly accessible document stores or social media profiles for such mapping. For example, remoteStorage can use WebFinger [27][28], a protocol designed for this purpose; the user identifier could include the user's Google Docs or Dropbox file URL that is publicly viewable, and is updated by the user to contain the contact token; or the user identifier could include the LinkedIn profile, and a predefined block or link on the profile page can contain the contact token. Such ideas delegate the storage of user reachability to external services.

Alternatively, a P2P network can be used to store the reachability data in a purely distributed fashion. Such a P2P network for call signaling has been proposed for VoIP/SIP based communication [29][30]. However, web apps cannot create listening sockets in JavaScript. Hence implementing such a P2P network for call signaling purely in web apps is not feasible without an external socket gateway, plugin, or an external P2P network. Such structured P2P networks or distributed hash tables (DHTs) can, however, be used for the control path network (replacing Fig.10) in a large-scale conference.

Exposing the FCM token to the public, in either the public data store, or a P2P network, has a risk that a malicious entity can send any push notification to the user's devices. A harmful notification may contain an attachment or a link to a malware. On the other hand, using a controlled notification server for push notification can restrict the content of the notification, and control the access to only the authorized senders.

*Serverless mode*

In the serverless mode, even though our app demonstrates the use of emails, the actual sending and receiving of the signaling messages is outside the scope of the app. More importantly, any other popular messaging service can be used. Using regular SMS is tricky due to the message size restrictions, in which case use of MMS with attachments containing the message is recommended.

A drawback of email based flow is that it is not real-time, i.e., there may be significant delay in sending and receiving the email. Hence, in this mode, the invite or accept message has an implicit expiration of two minutes. If the response is delayed by more than that, then the app may not use that message. Instant messaging, if available, is preferred to email messaging for exchanging the call signaling data.

Manual copy-paste of the call signaling message between the app and the messaging client may be cumbersome. Web applications support drag-and-drop. It should be possible to create draggable attachments containing the signaling message, which the user could drag-and-drop between this app and her messaging client. However, due to security concerns existing email clients (even those that are web based) do not allow such cross-app drag-and-drop for non-text and non-image content. To workaround, one could present the signaling message as an image, with the actual message text in the metadata content of the PNG image file, or as an image with QR code representing the signaling message, to allow drag-and-drop. We have not tested if this actually works.

For serverless offer-answer negotiation, a challenge is to determine how long the SDP offer and answer containing the ICE candidates are valid. This depends on the network devices near the endpoint, e.g., how long does the user's local router or NAT keep the port open if there is no traffic. In the absence of any NAT in the path, such as for connecting between endpoints within the same local area network, the offer and answer are valid for more than two minutes, beyond the call setup timeout of our app. However, many routers close an inactive port with timeout ranging from 30 seconds to 5 minutes. If the WebRTC implementation of the browser can refresh the STUN binding of the offer and answer internally, until the peer connection is successful or closed, then the web app does not need to worry about the validity of the open ports.

*Video conferencing web app*

Our client endpoint runs three separate web apps: Ezcall, wrapper and media-chat-peer. The purpose of Ezcall, that includes a service worker, is to do initial call signaling and notification; and the purpose of media-chat-peer web app is to provide the video conferencing user experience. The wrapper functionality, which glues the two, may be merged with the other two, reducing the number of web apps in the endpoint. However, we keep that separate so that the actual video conference web app can run in a controlled or sandboxed environment of the iframe if needed. This allows using other third-party apps, and the user can still impose certain security restrictions on what those apps can do.

The video conference app is in a window separate from the call signaling app. This allows the call signaling app to be resized to a small window, while showing a full size video conferencing window. It also supports messenger-style tabbed conversations, by keeping the conferences in separate browser tabs to support multiple conferences. It also works well with the desktop installed app (PWA). It fits well with the browser extension, where the main signaling app is launched as a small popup when the extension icon is clicked. This can be extended to a click-to-call button on third-party websites. On mobile, we use an iframe to load the wrapper of the conference app instead of a separate window, to avoid slowdown of background window.

As mentioned previously, we keep the peer connection of the initial control path separate from the peer connection of the media path. This may be optimized, e.g., by keeping a factory of peer connections, that is reused across different layers of the app. Creating the initial data channel, separate from the media streams keeps the message size of initial call signaling within the limit of what is allowed by push notifications. The subsequent negotiation messages can be sent on that data channel.

The web components of the RTC bricks project can also be optimized to allow peer connection reuse for multiple media streams. This will also benefit our project, not in full mesh media path, but when using SFU or MCU in the endpoint. Currently, the media-chat component implements only the full-mesh topology in the media path. It should be extended to include SFU and/or MCU in the endpoint. The video-io web component, in particular, should ensure that the video encoder is called only once when publishing the stream, even if there are multiple subscribers receiving the stream from this publisher, on separate peer connections.

*Auth token and security*

We use a simple auth token for the initial user lookup and call setup. The external signup process should ensure that the user identifier associated with the token belongs to the user, e.g., by sending the token by email during signup if an email address is used as an identifier. The auth token should be treated as a secret key by the user. The signup system should allow easily revoking any compromised tokens. A leaked token can allow identity theft on this app, where a malicious actor can pose as another user. The risk is similar to that of a leaked password, and additional security measures such as multifactor authentication or a single-signon using third-party authenticators can reduce the risk. We can use a client certificate during app install, and link the auth token with the certificate, e.g., as a fingerprint of the public key. In that case a leaked token does not compromise the security, as long the private key associated with the certificate is not leaked. The system then uses the client certificate for all secure connections (TLS) with the server, and also uses that for user identity verification at the server as well as at the other clients.

Our media-chat-peer web app reuses the registered user identifier, or email address (only the user part, not the domain), as the user's display name in the user interface. If a third-party web app is used, or when serverless call signaling is done, the display name can be impersonated by the conference user with undesired user experience. The accuracy of the display name in the conference is the responsibility of the particular conference app, and many existing web based conferencing apps do not impose a restriction. More important, however, is the association between the user identifier of the call signaling and that of the video conference app. If a third-party web app is used that has a separate user identification, then it becomes the responsibility of that web app to co-relate the two.

The web components of the RTC bricks project use the shared-storage component for maintaining the conference data structure. Currently, it has little or no access control, e.g., one user can update the data object created by another user, to alter the text message, or call state of others. A controlled web app works well, and co-operates with each other using such a shared-storage. But it is prone to misuse by a malicious app that gains access to the storage. This is similar to the Firebase real-time database, where the access control responsibility lies in the client app and the apps data access policies. Unlike this, for RTC bricks, a file system style access control is proposed. Such an access controlled shared-storage can ensure correct read and write permissions on the various resources created by different users. For example, the /sessions/aaa-bbb-ccc resource may be accessed only by that conference's users, and the sub-resource /users/alice is owned by that user, Alice, with full access, whereas other participants can only read that sub-resource; or the /messages sub-resource is append only, where any user can add a message object under it, but cannot modify an existing message object. We can reuse that access control in our PeerStorageImpl object to enable it on the distributed shared data storage.

In addition to the basic access control of the resources on the shared-storage, some data integrity and privacy may be desired for certain resources. For example, the text message resource may be signed by the sender's private key, so that others can verify that the message was indeed sent by that sender. Similarly, private messages sent from Alice to Bob, but stored as a resource on the shared storage, may be encrypted using Bob's public key, so that only Bob may decrypt and view the message. Many of these ideas are applicable to any web video conference app in general, and are outside the scope of our Ezcall app.

## 5. CONCLUSIONS AND FUTURE WORK

We have described a software architecture to implement an endpoint driven multiparty video conferencing app. We extend the centralized video conferencing app of the RTC bricks project to enable peer-to-peer network and data storage, and push notification based call signaling. Our effort includes a lightweight endpoint implementation with a call signaling app (Ezcall), a web video conference app (media-chat-peer), a wrapper app acting as a glue, and new web components (PeerNetworkImpl and PeerStorageImpl). All together, the new effort of the endpoint is in less than 3000 lines of code, mostly in JavaScript, and less than 300 lines of PHP code of the lightweight notification server.

Our app can be made serverless by using email messages to do call signaling. This avoids deploying or implementing a dedicated application/media server for video conferencing. Strictly speaking, it still uses email servers, but those are independent and ubiquitous, and not tied to our system. The lightweight nature of our software is particularly suitable for low cost communication scenarios, while preserving the user privacy unlike other big tech's "free" offerings.

Our software is still in its early stage. We mentioned a few limitations in the previous section. In addition to working on some of those limitations, we plan to continue exploring newer ways to innovate in the endpoint. This goes against the current trend of software-as-a-service (SaaS) in creating web apps, but is inline with the fundamental Internet principle of keeping the application logic and control in the endpoint. From the user's perspective, in the long term, owning the installed app without a subscription fee is cheaper than renting a SaaS app and paying forever.